\documentclass{article}
\title{Reply to cond-mat/0211660: Comments on "A model for fatigue in
ferroelectric perovskite thin films" published in Appl. Phys.
Lett, 76, 1060 (2000); addendum, ibid. p.3655 }
\author{Matthew Dawber and James F. Scott}
\date{Centre for Ferroics, Dept of Earth Sciences, Downing St,
Cambridge, CB2 3EQ, UK. \\ Email: mdaw00@esc.cam.ac.uk,
jsco99@esc.cam.ac.uk}
\begin{document}
\maketitle

Although the appropriate forum for discussion of published papers
should be the journal itself through the properly refereed process
where we are given an opportunity to have a reply published
simultaneously with the comments, Taganstev has chosen instead to
attack our paper through this unrefereed forum. In this context it
is worth noting that the comments that Tagantsev has made were in
fact submitted to Applied Physics Letters more than two years ago.
At that time we wrote a reply to these comments. The decision of
the referee at that time was that Tagantsev's  comment was wrong
and should not be published. Although we do not feel this is the
appropriate forum for this discussion, Tagantsev has chosen to
resume this debate in the unrefereed public forum of cond-mat and
so we feel the need to defend ourselves against the 8 points he
has raised.

Our model was a first attempt to produce a quantitative analytic
model for fatigue based on earlier work by Yoo and Desu, and as
such requires further testing and development. We note that our
model has already been extended and applied successfully by Wang
et al. (Physica Status Solidi A, \textbf{191} 482 (2002)).
Tagantsev's "model" for fatigue is untestable (falsifiable).We
would encourage feedback from other authors who have attempted to
apply our model, and are always happy to discuss our work. Please
contact us directly on the above email addresses if you have any
concerns about the publications mentioned here.

1. Taganstsev objects to our use of the Onsager expression for the
local field at an oxygen vacancy, preferring instead an expression
that is linear in the dielectric constant. One wonders what might
happen when a ferroelectric goes through its phase transition and
the dielectric constant (and hence local field, if one uses
Taganstev's expression) diverges. The Tagantsev model of
ferroelectric detonation in which internal fields diverge as a
ferroelectric material is field cooled through its transition
temperature does not seem to have been experimentally observed.
Perhaps one should look beyond undergraduate textbooks such as
Kittel's. A more detailed calculation of the effective charge on
an oxygen vacancy has been recently undertaken by Prof. S.A.
Prosandeev (cond-mat/0209019) in which he found that our result
was much more appropriate than Taganstev's.

2. Taganstev claims that our equation is quite different from that
of O'Dwyer. Simple inspection of the two equations show that this
is not true. In the high field limit sinh(x)=exp(x), and as our
local field is only 1.5 times the applied field Taganstev's claim
that use of this field changes the result by "orders of magnitude"
is clearly unfounded.

3.Tagantsev's point on equation 10 is taken. The reason that there
appears to be a change in the oxygen vacancy concentration at the
interface in the absence of an applied field is that we have used
the high field limit of the sinh(x) term in the diffusion equation
ie. (exp(x)). This means that our equations are not appropriate
for low fields, but it should be noted that during polarisation
switching high fields are applied. The reason for the use of the
exponential limit of the sinh term was so as to simplify the
derivation that followed.

4.We consider that the approximations we have taken are
appropriate for the situation. The applied field in our model is
very high because the applied potential falls across a quite
narrow depletion region in the ferroelectric. Therefore in our
opinion the space charge field is not significant compared to the
applied field.

Tagantsev is correct that one would expect to see an increase in
concentration of charge at the electrodes in non-ferroelectric
back-to-back Schottky diodes, however his claim that this has
never been observed is false. This has been observed for at least
twenty years in zinc oxide varistors. (e.g. Hayashi et al., J.
Appl. Phys. \textbf{58} 5754 (1982)) Thus his argument helps prove
our model - as confirmed by Hayashi.

5. The activation energy of electrons was used to calculate the
number of oxygen vacancies that would be charged, not the
concentration of oxygen vacancies. The activation energy of 0.7 eV
in fact corresponds to the trapping energy of Ti$^{3+}$ which is
known to be associated with oxygen vacancies. We originally
considered that the important activation energy originated from
the charge state of the oxygen vacancy. However following our new
ideas on oxygen vacancy ordering we believe that the entropy term
is more important than originally anticipated. We would refer
readers to further references on this subject for more details.
(J.F. Scott and M. Dawber, Appl. Phys. Lett. \textbf{76} 3801
(2000), M. Dawber and J. F. Scott, Integr. Ferroelectr.
\textbf{32} 951 (2001)  J. F. Scott, Ferroelectric Memories
(Springer, Heidelberg, 2000), pp. 134)

6. Ref. 6 of Tagantsev's paper was originally cited by us because
it gave a reasonable number for the depletion width, which we used
in our calculations. In the years since we published our paper we
have re-examined the data of reference 6. We are no longer
convinced that the current observed in this paper is in fact
Fowler-Nordheim tunneling. We do not wish to further criticize
this paper in this unrefereed forum, but any interested reader
should attempt to fit the data of ref 6 to a Schottky plot to see
the origin of our concerns. Readers can contact us directly for a
more detailed explanation of these concerns which are partly based
on our own unpublished results. We also have concerns about the
effective masses used in Tagantsev's analysis and the lack of
specification of carrier type (electron/hole). [Their m* = 1.4
m$_{e}$ (Bull Am Phys Soc, Seattle, March 2001) value disagrees by
x4 with the known electron band mass, and in undoped PZT films the
carriers are NOT holes.] We have previously discussed these
problems elsewhere (J.F. Scott, Integr. Ferroelectr. \textbf{42} 1
(2002).

7. In our original paper the figure was incorrectly labelled due
to technical error in journal production. Our addendum clearly
acknowledges this error. We apologize for any confusion caused by
this error. The prediction of the equations in both papers is in
line with the data of Mihara.

8. Our model does predict a frequency dependence for fatigue. This
has been seen in other papers than that of Colla et al., where it
is true that the use of different waveforms complicates
interpretation. Examples of such papers include, Lee et al, Appl.
Phys. Lett. \textbf{79} 821 (2001) ; Zhang et al., Ferroelectrics,
\textbf{259} 109 (2001).

\end{document}